\documentclass[prl,superscriptaddress,onecolumn]{revtex4}
\usepackage[ansinew]{inputenc}
\usepackage{graphicx}
\usepackage{amsmath}
\usepackage{amsthm}
\usepackage{bm}
\usepackage{layout}
\usepackage{float}
\usepackage{txfonts}
\usepackage{amsfonts}
\usepackage{amssymb}%
\newcommand{\ket}[1]{|#1\rangle}

\newcommand{\be}{\begin{eqnarray}}
\newcommand{\ee}{\end{eqnarray}}

\begin{document}
\title{The speed of quantum and classical learning for performing the $k$-th root of NOT}

\author{Daniel Manzano}
\email{manzano@ugr.es}
\affiliation{Departamento de F\'isica At\'omica, Molecular y Nuclear, Universidad de Granada, 18071 Granada,
Spain}
\affiliation{Instituto Carlos I de F\'isica Te\'orica y Computacional, Universidad de Granada,
18071 Granada, Spain}
\affiliation{Institute of
Quantum Optics and Quantum Information, Austrian Academy of
Sciences, Boltzmanngasse 3, A-1090 Vienna, Austria}

\author{Marcin Paw{\l}owski}
\email{dokmpa@univ.gda.pl}
\affiliation{Institute of Theoretical Physics and Astrophysics, University of Gda\'nsk, 80-952 Gda\'nsk, Poland}

\author{\v{C}aslav Brukner}
\email{caslav.brukner@univie.ac.at} \affiliation{Institute of
Quantum Optics and Quantum Information, Austrian Academy of
Sciences, Boltzmanngasse 3, A-1090 Vienna, Austria}
\affiliation{Faculty of Physics, University of Vienna,
Boltzmanngasse 5, A-1090 Vienna, Austria}

\begin{abstract}
We consider quantum learning machines -- quantum computers that
modify themselves in order to improve their performance in some way
-- that are trained to perform certain {\it classical} task, i.e. to
execute a function which takes classical bits as input and returns
classical bits as output. This allows a fair comparison between
learning efficiency of quantum and classical learning machine in
terms of the number of iterations required for completion of
learning. We find an explicit example of the task for which
numerical simulations show that quantum learning is faster than its
classical counterpart. The task is extraction of the $k$-th root of
NOT (NOT = logical negation), with $k=2^m$ and $m \in \mathbb{N}$.
The reason for this speed-up is that classical machine requires
memory of size $\log k=m$ to accomplish the learning, while the
memory of a {\it single} qubit is sufficient for the quantum machine
for any $k$.
\end{abstract}

\maketitle

Learning can be defined as the changes in a system that result in an
improved performance over time on tasks that are similar to those
performed in the system's previous history. Although learning is
often thought of as a property associated with living things,
machines or computers are also able to modify their own algorithms
as a result of training experiences. This is the main subject of the
broad field of ``machine learning''. Recent progress in quantum
communication and quantum computation~\cite{nielsenchuang} --
development of novel and efficient ways to process information on
the basis of laws of quantum theory -- provides motivations to
generalize the theory of machine learning into the quantum
domain~\cite{brassard}. For example, quantum learning algorithms
have been developed for extracting information from a ``black-box''
oracle for an unknown Boolean function~\cite{alp,hunziger}.

The main ingredient of the quantum machine is a feed-back system
that is capable of modifying its initial quantum algorithm in
response to interaction with a ``teacher'' such that it yields
better approximations to the intended quantum algorithm. In the
literature there have been intensive and extensive studies by
employing feed-back systems. They include quantum neural
networks~\cite{qnn}, estimation of quantum states~\cite{qse}, and
automatic engineering of quantum states of molecules or light with a
genetic algorithm~\cite{laser,fempto,chemistry}. Quantum neural
networks deal with many-body quantum systems and refer to the class
of neural network models which explicitly use concepts from quantum
computing to simulate biological neural networks~\cite{qnn2}.
Standard state-engineering schemes optimize unitary transformations
to produce a given target quantum state. The present approach of
quantum automatic control contrasts with these methods. Instead of
quantum state it optimizes {\it quantum operations (e.g. unitary
transformations) to perform a given quantum information task}. It is
also different than the problems studied in
Ref.~\cite{alp,hunziger}, where one does not learn a task but rather
a specific property of a black-box oracle.

An interesting question arises in this context: (1) Can a quantum
machine learn to perform a given quantum algorithm? This question
has been answered affirmative for special tasks, such as quantum
pattern recognition~\cite{qpattern}, matching of unknown quantum
states~\cite{qmatching}, and for learning quantum computational
algorithms such as the Deutch algorithm~\cite{BAN}, the Grover
search algorithm and the discrete Fourier transform~\cite{GAM}.

Another interesting question is: (2) Can one have quantum
improvements in the speed of learning in a sense that a quantum
machine requires {\it fewer} steps than the best classical machine
to learn some {\it classical} task? By ``classical task'' we mean an
operation or a function which has classical input and classical
output. Quantum machines such as quantum state discriminator,
universal quantum cloner or programmable quantum
processor~\cite{buzekhillery} do not fall into this category.
Quantum computational algorithms do perform classical tasks, but no
investigation has been undertaken to compare speed of learning of
these algorithms with that of their classical counterparts. To our
knowledge the question (2) is still open thus far. In this letter we
will give evidence for the first explicit classical computational
task that quantum machines can learn {\it faster} than their
classical counterparts. In both cases certain set of independent 
parameters must be optimized to learn the task. We will show that 
the fraction of the space of parameters, which correspond to 
(approximate) successful completion of the task, is exponentially 
smaller for the classical machine than for the quantum one. This
analytical results supports our numerical simulation showing that
quantum machine learns faster than the classical one.

We first define a family of problems of our interest:
let $m$-th member ($m \in \mathbb{N}$) of this family be the $k$-th 
root of NOT with $k=2^m$, where the roots of NOT are defined as follows:

\vspace{5mm}
$\textrm{\bf{Definition 1.}}$ The operation is $k${\it-th root of NOT}  if,
when applied subsequently $nk$ times on the Boolean input of 0 or 1,
it returns the input for even $n$'s and its negation for odd $n$'s.
We denote this operation with $\sqrt[k]{\mbox{NOT}}$. (Remark: With
this definition we want to discard the cases for which, for example,
the operation returns $k$-th root of NOT when performed once, but
does not return identity when performed twice.)

The machine that performs this operation takes one input bit and
returns one output bit. This bit will be called ``target bit''. In
general, however, the machine could use many more auxiliary bits
that might help the performance. Specifically, in the classical case
the input $\vec{i}$ and output $\vec{j}$ are vectors with binary
components. Any operation is defined by a probability distribution
$p(\vec{i},\vec{j})$ which gives the probability that the machine
will generate the output $\vec{j}$ from the input $\vec{i}$. Thus,
one has $\sum_{\vec{j}} p(\vec{i},\vec{j}) =1$. The readout of the
target bit is a map: $\vec{j} \to \{0,1\}$. Without loss of
generality we assume that the target bit is the first component of
the input and the output vector. The remaining components are
auxiliary bits which play the role of the machine's memory.

In quantum case no auxiliary (qu)bits are necessary as only one
qubit is enough to implement any $\sqrt[k]{\mbox{NOT}}$. The input
of the machine is a single qubit and the machine itself is a unitary
transformation. The input state will be either $\ket{0}$ or
$\ket{1}$ corresponding to the Boolean values of classical bits
``0'' and ``1'', respectively. The readout procedure is the
measurement in the computational basis $\{ \ket{0},\ket{1} \}$ and
we consider the state that the qubit is projected to as the output
of the machine.

In both cases the term learning is used for the process of approximating the
 function $\sqrt[k]{\mbox{NOT}}$ to which we will refer as the target
function. We will consider that learning has been accomplished when the learning machine
returns with high probability correct outputs for both inputs. Then a learning 
process is reduced to approximating the target function
in a sequence of taking the inputs, performing transformations on the inputs, returning the
outputs, estimating the fidelity between the actual outputs and the ones that 
the target function would have produced and correspondingly
of making adjustments to the transformations. The schematic diagrams depicting both
types of machines are shown in the Fig. 1. Now we will describe the learning in both cases
in more detail.

\textit{Quantum learning}: In every learning trial the following
steps are performed:

\begin{enumerate}
\item Select a new unitary operator $U$ using a Gaussian random walk
(The first $U$ is initialized randomly using the Haar measure).

\item Run the unitary $U^k$ on an input qubit state chosen to be
$|0\rangle$ or $|1\rangle$ with equal probability. Measure the
output qubit in the computation basis. Repeat this on $M$ input
states and store the results (classical bits). The number $M$
defines the size of teachers (classical) memory of the quantum
machine.

\item Estimate how close is the actual operation to the target one. To
achieve this count the number of times the operation is successful
in approximating the target function (i.e. it produces $|1\rangle$,
when the input was in $|0\rangle$, and it produces $|0\rangle$, when
it was in $|1\rangle$). The number of successes is denoted by
$new_s$ and $old_s$ in the executed and the previous trial,
respectively.

\item If $new_s \ge old_s$, go to 1 with the current unitary operator
as the center of the Gaussian; Otherwise, go to 1 with the unitary
operator chosen in the previous trail as the center of the Gaussian.
\end{enumerate}

\begin{figure}[t]
\includegraphics[scale=0.45]{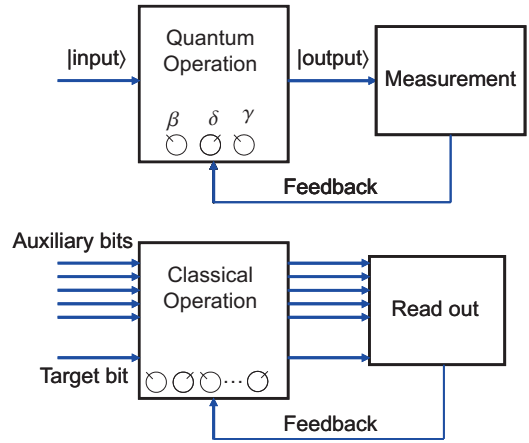}
\caption{Diagram of classical and quantum learning machines. The
learning procedure consists in a sequence of taking the inputs,
performing transformations on them, returning the outputs,
estimating the figure of merit between the outputs obtained and the
expected ones and correspondingly making adjustments to the
transformations. For the task of extracting the $k$-th root of NOT
(see text for definition) the dimension of the space of 
parameters for a classical machine is log 2k larger than that 
for a quantum machine.}
\end{figure}

Any single qubit rotation can be
parameterized by Euler's angles as follows:
\begin{equation}\label{eq:parameter1}
U=e^{i\alpha} \left(
\begin{array}{cc}
e^{-i\left(\frac{\beta}{2}+\frac{\delta}{2}\right)}\cos\left(\frac{\gamma}{2}\right) & -e^{i\left(-\frac{\beta}{2}+\frac{\delta}{2}\right)}\sin\left(\frac{\gamma}{2}\right)\\
e^{i\left(\frac{\beta}{2}-\frac{\delta}{2}\right)}\sin\left(\frac{\gamma}{2}\right)
&
e^{i\left(\frac{\beta}{2}+\frac{\delta}{2}\right)}\cos\left(\frac{\gamma}{2}\right) \\
\end{array}
\right).
\end{equation}
Since the global phase $\alpha$ is irrelevant for the present
application, we are left with the parameters $\delta\in [0,2\pi]$,
$\beta\in [0,2\pi]$, $\gamma\in [0,\pi]$. In every new learning
trial these parameters will be selected independently with a normal
probability distribution centered around the values from the
previous run and the widths of the Gaussians are taken as free
parameters of the simulation. There are two free parameters of the
learning procedure: $\sigma_\gamma$ and $\sigma_\beta$
($\sigma_\delta=\sigma_\beta$). In all simulations these parameters
are optimized to minimize the number of learning steps.

Note that if quantum machine performs the task for $n=1$ perfectly,
then it will also perform the task perfectly for all $n$. This is
why our quantum machine is trained only to learn the task for $n=1$.
Nevertheless, after the learning has been completed one should
compare how close the performance of the learning machine is to this
of the target operation for all $n$. We define a set of figures of
merit $\left\{ P^n \right\}_{n=1}^{\infty}$ as follows:
\begin{eqnarray}
P^1&=&\frac{1}{2} ( | \langle 0|U^k|1 \rangle |^2 + | \langle 1|U^k|0 \rangle |^2 ) \label{fidelity} \\
P^2&=&\frac{1}{4} ( | \langle 0|U^k|1 \rangle |^2 + | \langle 1|U^k|0 \rangle |^2 + | \langle 0|U^{2k}|0 \rangle |^2 + \langle 1|U^{2k}|1\rangle |^2 ) \nonumber \\
P^n&=&\frac{1}{2n}(| \langle 0|U^k|1 \rangle |^2 + | \langle 1|U^k|0
\rangle |^2 + \nonumber \\ & & \dots +  |\langle 0|U^{nk}|b\rangle
|^2 + |\langle 1|U^{nk}|b\oplus1\rangle |^2), \nonumber
\end{eqnarray}
where $b=0$ if $n$ is even, and $b=1$ if $n$ is odd, and $\oplus$
denotes sum modulo 2. Note that each subsequent $P^n$ is more
demanding in the sense that more constraints from the definition of
the $\sqrt[k]{\mbox{NOT}}$ are being taken into account. This is
reflected by the resultsprocedure, which are presented in Fig. 2.

The memory size of the teacher $M$ is another free parameter of the
quantum machine. The learning ability has a very strong dependence
on $M$ as can be seen from Fig. 3. For lower values of M $M$ the
learning is faster at the beginning (up to about 4x$10^4$ trials),
before it slows down and saturates. At the saturation the size of
the memory does not allow distinguishing between sufficiently
``good'' operations all for which $new_s = M$. For higher $M$ values
the learning is slower, but it reaches higher fidelities. To combine
the high speed with the high fidelity of learning we apply the
learning procedure with variable $M$: The machine starts with $M=1$
and whenever it obtains the number of successes $new_s=M$ it
increments $M$ by one. With this kind of algorithm the learning has
one less free parameter. All our simulations were done for variable
$M$, unless stated otherwise.

\begin{figure}[t]
\begin{center}
\hspace{-1.2cm}
\includegraphics[scale=0.5]{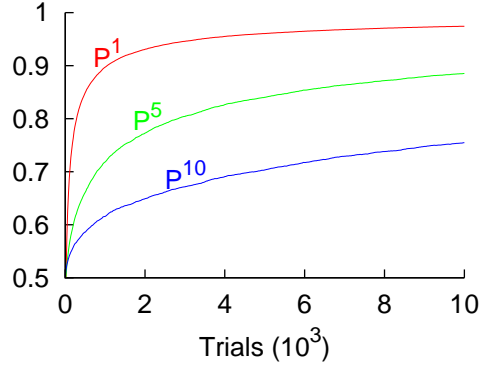}
\end{center}
\vspace{-1cm} \caption{Quantum learning for performing the $4$-th
root of NOT. Different figures of merit $P^s$ ($s=1,5,10$) as a
function of the number of learning trials (x $10^3$). The size of
teachers memory $M$ is varied to achieve the maximal value of the
figures of merits for a given number of trials. The free parameters
have the values $\sigma_\gamma=\frac{\pi}{4}$ and
$\sigma_\alpha=\sigma_\beta=\frac{\pi}{8}$.}
\end{figure}

\begin{figure}[t]
\begin{center}
\hspace{-1.2cm}
\includegraphics[scale=0.5]{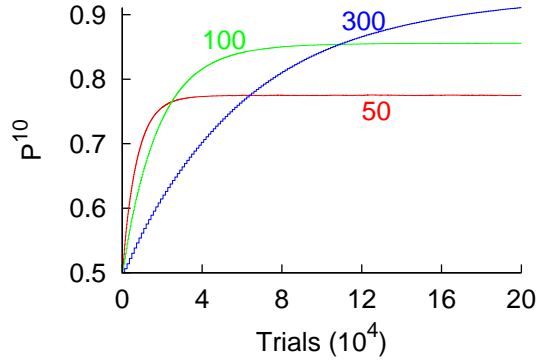}
\end{center}
\vspace{-1cm} \caption{Quantum learning for performing the $4$-th
root of NOT. Figure of merit $P^{10}$ as a function of the number of
learning trials (x $10^4)$ for different sizes of teachers memory
$M$ (blue = 300, green =100, red=50). The free parameters
have the values $\sigma_\gamma=\frac{\pi}{4}$ and
$\sigma_\alpha=\sigma_\beta=\frac{\pi}{8}$}
\end{figure}

Next we describe the classical learning procedure.

\textit{Classical learning}: The classical learning is an iterative
process of finding the optimal probability distribution
$p(\vec{i},\vec{j})$ for the classical machine to extract the
$\sqrt[k]{\mbox{NOT}}$. The speed of learning depends on the number
$N^2-N$ of independent parameters (independent probabilities
$p(\vec{i},\vec{j})$), where $N=2^{dim(i)}$ and $dim(i)$ is the
dimension of the input $\vec{i}$ and the output vector $\vec{j}$. We
will refer to $N$ as the memory size of the classical learning
machines because it is equal to the total number of distinguishable
internal states of the machine. To minimize the number of learning
trials required to complete the learning and thus to maximize the
speed of learning, we are interested in the minimal number of
internal states $N$ for which it is possible to construct a
classical machine that is able to extract $\sqrt[k]{\mbox{NOT}}$.

\vspace{5mm}
$\textrm{\bf{Lemma}}$: Any classical machine that performs $k$-th
 root of NOT perfectly must have at least $2k$ internal states if $k=2^m$
 and $m\in \mathbb{N}$.

\vspace{5mm}
$\textrm{\bf{Proof. }}$ Each probabilistic classical machine can be
considered as a convex combination of deterministic ones. If it
performs some task perfectly, then there must also be deterministic
machine that does the same. This means that we can restrict
ourselves in this proof only to deterministic machines without any
loss of generality. Any (deterministic, classical) machine can be
represented as an oriented graph, with vertices corresponding to the
internal states. Edge pointing from vertex $\vec{i}$ to $\vec{j}$
will mean that the operation on input $\vec{i}$ generates the output
$\vec{j}$. Any (finite) machine must have at least one loop and, if
the machine is run subsequently a large number of times, it will
eventually end up in that loop. Since the definition of the task
involves arbitrary large $n$'s we may start our analysis from $n$
large enough such that the machine is already in the loop. Since we
will prove the lemma by giving constraint from below on the size of
the loop, we may assume that the whole graph is a one loop and each
vertex is a part of it.

Let the length of the loop be $N$. Let $g$ be the greatest common
divisor $g=GCD(k,N)$. Then there exist numbers $x$ and $y$ such that
\be k=gx \quad N=gy \quad GCD(x,y)=1 \ee If the machine is initially
in a vertex that corresponds to input ``1'' of the target bit and we
apply the operation $Nk$ times we will always end up in the same
vertex ``1'', since $N k=0 \mod N$. Since, however, the task is
defined such that for $N$ odd the ending vertex should correspond to
``0'' value for the target bit, one concludes that $N$ must be even.
Therefore, we can write $N$ as $N=2^K c$, where $c$ is odd and $K\ge 1$. 
We also have
$Nx=0\mod N$, but since $Nx=gyx=ky$, then $ky=0 \mod N$. According
to our definition of $\sqrt[k]{\mbox{NOT}}$ this implies that $y$ is
even and, since $GCD(y,x)=1$, $x$ is odd. Also \be
y=\frac{N}{g}=\frac{Nx}{k}=2^{K-m}xc \ee Since $y$ is even and both
$x$ and $c$ odd, then $K\ge m+1$ must hold. We conclude with
$N=2^Kc\ge 2^K\ge 2^{m+1}=2k$. 

\vspace{5mm}
The Lemma implies that if the machine is to perform
$\sqrt[k]{\mbox{NOT}}$ perfectly it needs to have $\log k=m$
auxiliary bits in addition to the target bit. It is easy to check
that this is not only necessary but also a sufficient condition. One
just needs to design machine that is a loop of length $2k$ where the
vertices corresponding to initial target input bits 0 and 1 are at a
distance $k$ from each other. The number of functions with this
property divided by the total number of functions $f:
\{0,1\}^{2k}\to\{0,1\}^{2k}$ gives the fraction of the target functions:
\begin{equation}\label{eq:ratio}
R=\frac{(2k-4)!(2k-2)(k^2-2)}{(2k)^{2k}} \simeq O\left(\frac{1}{k
4^k}\right).
\end{equation}
The target functions thus constitute an exponential small fraction
of all functions. Next, we will consider probabilistic classical
machines which in order to approximate the target functions with
high probability need to be sufficiently ``close'' (e.g. in the
sense of Kullback-Leibler divergence) in the probability space. In
such a way both the quantum and classical machines ``search'' in a
continuous space of parameters, however, the relative fraction of
this space that is close to the target functions is obviously much
larger for the quantum case.

In the case of quantum machine any root of NOT can be performed with
only one qubit. The operation that performs $\sqrt[k]{\mbox{NOT}}$
is $\sqrt[k] {\sigma_x}$, where $\sigma_x$ is spin matrix along
direction $x$. Therefore, the memory requirements for our family of
problems grows as $\log k$ in the classical case, while remaining
constant in the quantum one.

Next, we introduce the classical learning procedure. We assume that
the classical machine is initially in a ``random'' state for which $
p(\vec{i},\vec{j})=\frac{1}{2k}.$ The learning process consist of
the following steps:

\begin{enumerate}

\item Set initially the internal state of the machine such that its
first bit (target bit) is in 0 or 1 with equal probability. All
auxiliary bits are in 0.

\item Apply the operation $k$ times and after each of them read out the
output: $\vec{j}_r$, with $r \in \{1,...,k\}$. We observe a sequence
$\vec{i} \equiv \vec{j}_0 \rightarrow \vec{j}_1 \rightarrow
\vec{j}_2 \rightarrow ... \rightarrow \vec{j}_{k}$ of machine'
states. If the target bit of the final state $\vec{j}_k$ is inverse
of the target bit of initial state $\vec{i}$, move to step 3.
Otherwise move to 4.

\item Increase every probability $p(\vec{j}_{r-1} ,\vec{j}_r)$ that led
to success by adding a factor $1 \geq K_s \geq 0 $. Renormalize the
probability distribution such that $\sum_{\vec{j}}
p(\vec{i},\vec{j})=1$ and go back to step 2.

\item Decrease every probability $p(\vec{j}_{r-1} ,\vec{j}_r)$ that led
to a failure by subtracting a factor $1 \geq K_f \geq 0$ (if then
the probability is negative, put it to be 0). Renormalize the
probability distribution and go back to step 1.
\end{enumerate}

Note that repeating the steps 2. and 3. the classical machine
gradually learns to perform the task for all $n$. The learning has
two free parameters $K_s$ and $K_f$, exactly like the quantum
learning (with a variable teachers memory size $M$). To estimate how
close is machine's functioning to the one of the target machine we
use the set of figures of merit for all $n$: $\left\{ P^n
\right\}_{n=1}^{\infty}$, which are similar to those of
Eq.~(\ref{fidelity}). For example, $P^2 \equiv P_k(0,1)+ P_k(1,0) +
P_{2k}(0,0) + P_{2k}(1,1)$, where $P_k(1,0)$ is the probability that
the target bit has been changed from 1 to 0 after applying the
transformation $k$ times and other probabilities are similarly
defined.

\begin{figure}
\hspace{-1.2cm}
\includegraphics[scale=0.52]{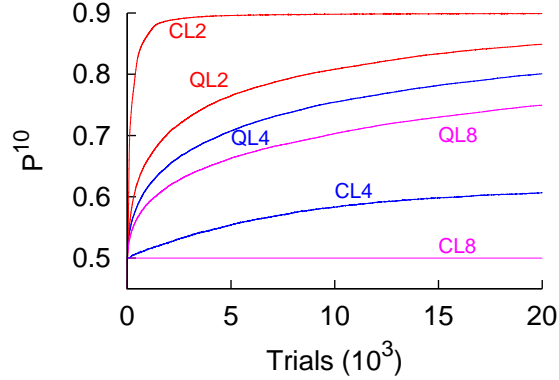}
\vspace{-0.5cm} \caption{The figure of merit ($P^{10}$) of classical
(CL) and quantum (QL) learning for performing different $k$-th
($k=2,4,8$) roots of NOT as a function of the number of learning
trials (x$10^3$). The values of free parameters are chosen to
maximize the figure of merit. Already for $k=4$ quantum learning is
faster than the classical one. For the 8-th root of NOT, the figure
of merit of classical learning is as for a random choice ($=0.5$) at
the given time scale. The free parameters have the values
$\sigma_\gamma=\frac{\pi}{4}$ and
$\sigma_\alpha=\sigma_\beta=\frac{\pi}{8}$ (for all roots)  for the
quantum case and the values $K_s=K_f=0.25$ (2nd root),
$K_s=K_f=0.75$ (4th root) and $K_s=0.75\;K_f=0.25$ (8th root) for
the classical one.}
\end{figure}

We have performed computer simulations of the both quantum and
classical learning process. The results are presented in Fig. 4. We
see that the learning in the quantum case is much faster for $k>2$.
This speed-up can be understood if one realizes that for the present
 problem the process of learning is an optimization of a square matrix:
unitary transformation $U$ in the quantum case, and a matrix with
entries $p(\vec{i},\vec{j})$ in the classical one. While the size of
$U$ remains 2 (with complex entries), the size of the matrix with
entries $p(\vec{i},\vec{j})$ grows linearly with $k$. It is clear
that optimization of significantly larger matrices requires more
iterative steps and thus leads to slower learning.

The classical learning algorithm given is not the most general and
might not be optimal. The general framework for finding optimal
learning procedures is still not fully understood. We have chosen
the quantum and classical learning algorithms such that the
comparison between them is most evident. The two tasks, i.e. finding
a unitary operator for the $k$-th root of NOT, and finding a
classical probability distribution that generates the $k$-th root of
NOT, though are different from the physical point of view, both
require optimization of matrix elements. Since for a given task, the
classical machines require a significantly larger number of
independent parameters (of which only a small fraction leads to the
desired matrix) to be optimized, it is natural to assume that they
also require a larger number of learning steps to accomplish
learning, regardless of the explicit learning procedure employed.
This is exactly what our numerical simulations show.

Quantum information processing has been shown to allow a speed-up
over the best possible classical algorithms in computation and has
advantages over its classical counterpart in communication tasks,
such as secure transmission of information or communication
complexity. In this paper we extend the list with a novel task from
the field of machine learning: learning to perform the $k$-th root
of NOT.

\vspace{10mm}
$\textrm{\bf{Acknowledgments}}$

\vspace{3mm}
We acknowledge support from the EC Project QAP (No. 015848), the
Austrian Science Foundation FWF within Projects No. P19570-N16, SFB
and CoQuS No. W1210-N16, the Ministerio de Ciencia e Innovaci\'on
(Fellowship BES-2006-13234) and the Instituto Carlos I for the use
of computational resources. The collaboration is a part of an
\"{O}AD/MNiSW program. {\v C}.B. thanks J. Lee for discussions.


\begin{thebibliography}{99}
\bibitem{nielsenchuang} M. A. Nielsen and I. L. Chuang, Quantum Computation
and Quantum Information (Cambridge University Press, 2000, UK).
\bibitem{brassard} E. Aimeur, G. Brassard, S. Gambs, {\it Canadian Conference on AI 2006} pp 431-442.
\bibitem{alp} A. Atici, and R.A. Servedio, Quant. Inf. Proc., {\bf 4}, 355, (2005).
\bibitem{hunziger} M. Hunziker, D.A. Meyer, J. Park,J. Pommersheim, and M. Rothstein, arXiv:quant-ph/0309059.
\bibitem{qnn} G.
Purushothaman and N. B. Karayiannis, IEEE Trans. Neural Networks,
\textbf{8}:679 (1997).
\bibitem{qnn2} E.C. Behrman, L.R. Nash, J.E. Steck, V.G. Chandrashekar and S.R. Skinner. Information Sciences, \textbf{128}:257 (2000).
\bibitem{qse} D. G. Fischer, S. H. Kienle, and M. Freyberger, Phys. Rev. A, \textbf{61}:032306, 2000.
\bibitem{laser} R. S. Judson and H. Rabitz, Phys. Rev. Lett., \textbf{68}:1500 (1991).
\bibitem{fempto} T. Baumert, T. Brixner, V. Seyfried, M. Strehle, and G. Gerber, Appl. Phys. B, \textbf{65}:779 (1997).
\bibitem{chemistry} A. Assion, T. Baumert, M. Bergt, T. Brixner, B. Kiefer, V. Seyfried, M. Strehle, and G. Gerber, Science \textbf{282}:919 (1998).
\bibitem{qpattern} R. Neigovzen, J. Neves, R. Sollacher and S. J. Glaser,  Phys. Rev. A \textbf{79}, 042321 (2009).
\bibitem{qmatching} M. Sasaki and A. Carlini, Phys. Rev. A \textbf{66}, 22303 (2002).
\bibitem{BAN} J. Bang, J. Lim, M.S. Kim and J. Lee, arXiv:0803.2976.
\bibitem{GAM} S. Gammelmark and K. Molmer, New J. Phys. \textbf{11}, 033017 (2009).
\bibitem{buzekhillery} M. Hillery, and V. Buzek,  Contemporary Physics {\bf 50}, 575-586 (2009).

\end{thebibliography}
\end{document}